\documentclass{article}
\usepackage{graphicx}
\usepackage{fancyhdr}
\usepackage{setspace}

\usepackage[paper=a4paper,left=35mm,right=35mm,top=30mm,bottom=30mm, twoside]{geometry}   

\begin{document}


\begin{singlespace}

\title{Atomic-scale Oxygen-mediated Etching of 2D MoS$_2$ and MoTe$_2$}

\maketitle





\noindent E. H. {\AA}hlgren, A. Markevich, S. Scharinger, B. Fickl, G. Zagler, F. Herterich, C. Mangler, J. Kotakoski\\
    {\it Faculty of Physics, University of Vienna, 1090 Vienna, Austria\\
    Email Address: harriet.ahlgren@univie.ac.at}\\

\noindent
N. McEvoy\\
    {\it AMBER \& School of Chemistry, Trinity College Dublin, Dublin 2, Ireland}\\



\end{singlespace}

\begin{abstract}

Oxidation is the main cause of degradation of many two-dimensional materials, including transition metal dichalcogenides, under ambient conditions.
Some of the materials are more affected by oxidation than others. To elucidate the oxidation-induced degradation mechanisms in transition metal dichalcogenides, the chemical effects in single layer MoS$_2$ and MoTe$_2$ were studied in situ in an electron microscope under controlled low-pressure oxygen environments at room temperature.
MoTe$_2$ is found to be reactive to oxygen, leading to significant degradation above a pressure of 1$\times$10$^{-7}$~torr.
Curiously, the common hydrocarbon contamination found on practically all surfaces accelerates the damage rate significantly, by up to a factor of forty.
In contrast to MoTe$_2$, MoS$_2$ is found to be inert under oxygen environment, with all observed structural changes being caused by electron irradiation only, leading to well-defined pores with high proportion of molybdenum nanowire-terminated edges.
Using density functional theory calculations, a further atomic-scale mechanism leading to the observed oxygen-related degradation in MoTe$_2$ is proposed and the role of the carbon in the etching is explored.
Together, the results provide an important insight into the oxygen-related deterioration of two-dimensional materials under ambient conditions relevant in many fields.

\end{abstract}

\newpage


\section{Introduction}
Two-dimensional (2D) transition metal dichalcogenides (TMDs) have received significant attention recently due to their possible applications in optoelectronics~\cite{singh2019-opto-review}, catalysis~\cite{Aguilar2020,Li2020}, gas sensing~\cite{Ramu2019}, medical sensing~\cite{Liu2019} and nanoelectronics~\cite{Larentis2017}.
This research has revealed that while some of them are inert to atmospheric changes for extended periods of time, others corrode within hours after production.
For example, 2D MoS$_2$ is often considered inert under ambient conditions, its properties deteriorating only after several weeks of ambient exposure~\cite{Chamlagain2021} or when immersed in water~\cite{Zhang2017}.
In contrast, MoTe$_2$ has been reported to be one of the most reactive TMDs~\cite{Mirabelli2016}.
However, not much is known about the atomic-level processes leading to the drastically different behavior of these materials.

Aberration corrected (scanning) transmission electron microscopy [(S)TEM] provides access to the exact atomic structure of materials with a sub-second time resolution.
However, high energy electrons used for imaging can also cause structural changes, as has been already demonstrated for both MoS$_2$ and MoTe$_2$.
In MoS$_2$, continuous electron exposure leads quickly to the formation of sulfur vacancies~\cite{Komsa2013} through a combined effect of electronic excitations and knock-on damage~\cite{Kretschmer2020}, which first agglomerate into vacancy lines before pores with molybdenum-rich edges appear~\cite{Liu2013}.
In contrast, presumably due to suppressed knock-on damage resulting from the larger mass of Te compared to S, vacancy formation in MoTe$_2$ is considerably slower allowing dynamic phase transitions to take place without removal of atoms~\cite{elibol2018-mote2}.
Nevertheless, structural changes are inevitable during prolonged imaging of both materials.
Therefore, in order to study oxidation-related structural changes, they must be separated from pure electron-irradiation-induced effects.
This requires an instrument with ultra high vacuum and means to introduce a controlled low-pressure atmosphere around the sample during imaging~\cite{hotz_ultra-high_2016}.
Such experiments have already revealed that chemical etching in graphene takes place at a partial oxygen pressures of $>3\times 10^{-8}$~torr~\cite{leuthner_chemistry_2021}, well below typical pressures of TEM instruments with side entry holders, leading to pore growth that starts at defective sites~\cite{Leuthner2019}. Pristine graphene areas remain unaffected.
However, similar studies remain lacking for all other 2D materials.

Here, we use the same strategy to compare the behavior of suspended 2D MoS$_2$ and MoTe$_2$ mono-layers under low-pressure ($9\times 10^{-10} - 4\times 10^{-7}$~torr) oxygen atmospheres in situ while being imaged at atomic resolution via STEM. 
Under electron irradiation O$_2$ molecules can split into atomic oxygen, accelerating the chemical effects up to an experimentally accessible time scale.
In our experiments, structural damage in MoS$_2$ shows no dependency on the oxygen partial pressure, displaying the well-known~\cite{Komsa2013,Liu2013,Huang2018} electron-beam-related creation of vacancies and later pores with molybdenum-rich edge structures.
In contrast, in MoTe$_2$ there is a marked difference in structural changes at different oxygen pressures.
Specifically, in ultra high vacuum, damage in MoTe$_2$ is similar to that in MoS$_2$ apart from the lack of formation of Mo nanowires at the pore edges, but an increase in the oxygen level leads to increased etching damage rate.
Surprisingly, it also depends on the amount of local hydrocarbon surface contamination, which can lead to an additional increase by up to a factor of 40.
Our ab initio simulations reveal a marked difference between atomic oxygen binding in the two materials, revealing a possible etching mechanism for MoTe$_2$ involving O adsorption on neighboring Te sites that leads to substitution of one Te with an oxygen atom and desorption of a weakly bound TeO molecule (binding energy of ca. $0.21$~eV). Carbon adsorbed on the structure further decreases the binding energy.
These results provide for the first time direct atomic-level insight into oxygen-related structural changes in TMDs and explain the drastic difference between the behavior of MoS$_2$ and MoTe$_2$ under extended exposure to atmospheres containing oxygen.

\section{Results and discussion}

The MoS$_2$ samples were grown via chemical vapor deposition~\cite{O'Brien2014} on SiO$_2$, whereas the MoTe$_2$ samples were prepared on a similar substrate via mechanical exfoliation using adhesive tape.
In both cases, they were transferred onto holey carbon support grids without the use of polymer, and baked at ca. 150$^\circ$C in vacuum before being inserted into the Nion UltraSTEM 100 microscope.
All images were recorded at 60~kV acceleration voltage with a typical beam current of ca. $25$~pA.
The angular range for the high angle annular dark field (HAADF) detector was $80-300$~mrad, and the convergence semi-angle of the electron probe was $30$~mrad.

\begin{figure}
\begin{center}
\includegraphics[width=1.0\textwidth]{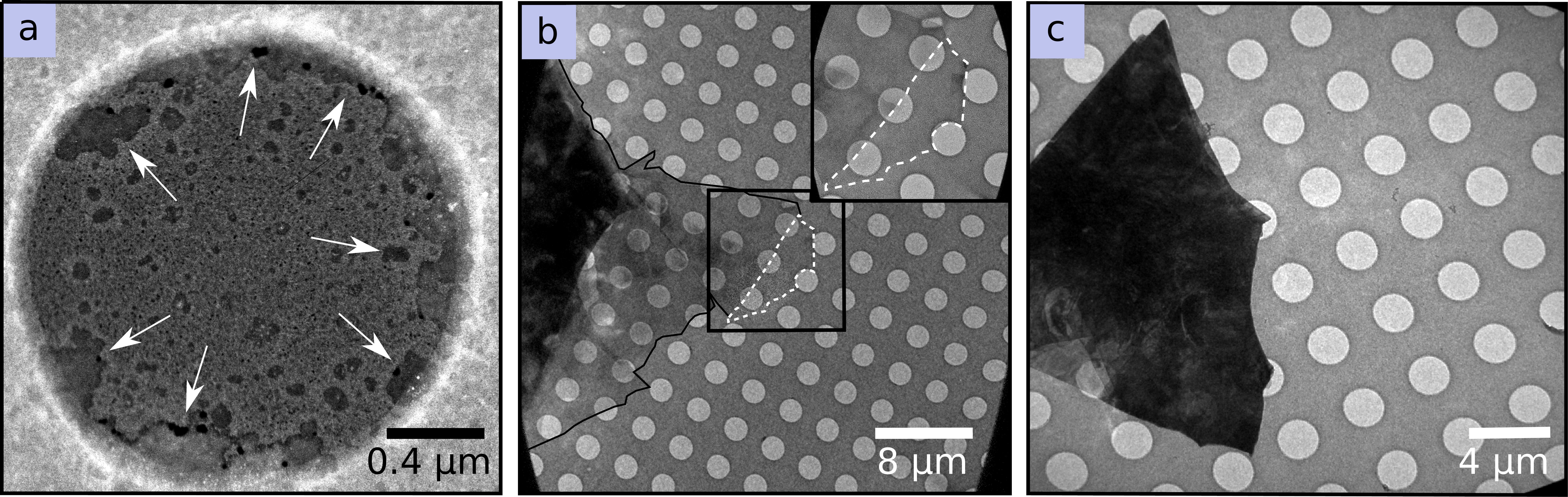}
    \caption {
        {\bf Effects of air exposure on MoTe$_2$.}
        (a) A STEM-HAADF image of a few-layer MoTe$_2$ flake suspended over a hole in amorphous carbon film and exposed to air for approximately one day.
        Some examples of severely damaged sample areas are highlighted with the arrows.
        (b) A bright field Ronchigram image of a pristine MoTe$_2$ sample with various local thicknesses on amorphous holey carbon film.
        The sample edge is outlined with a back continuous line, and the mono-layer area is indicated with a white dotted line.
        A further magnification of the thin area is shown as an inset.
        (c) The same sample area as in panel b after a $6$~h air exposure.
        Only the thicker area remains on the film while the mono-layer and few-layer areas have disappeared.}
\label{fg::air-exposure}
\end{center}
\end{figure}

While the MoS$_2$ samples did not show clear effects to air exposure over several months, for MoTe$_2$ even a relatively brief air exposure led to sample degradation.
This can be clearly seen in \textbf{Figure~\ref{fg::air-exposure}}a, displaying a STEM-HAADF image of a suspended MoTe$_2$ sample area that had been exposed to air for approximately 24 hours.
Overall on the sample, the 2D crystalline film has turned into a porous structure.
Another example is shown in Figure~\ref{fg::air-exposure}b-c displaying larger scale bright field images (recorded with the Ronchigram camera in the same instrument) before (panel b) and after (panel c) six hours air exposure.
All of the thin area within the sample (marked by the black and white lines in panel b) has completely disappeared during exposure to air.
We note that similar damage was shown in Ref.~\cite{elibol2018-mote2}, however, without connecting it to the ambient exposure of the sample.

\begin{figure}
\begin{center}
\includegraphics[width=1.0\textwidth]{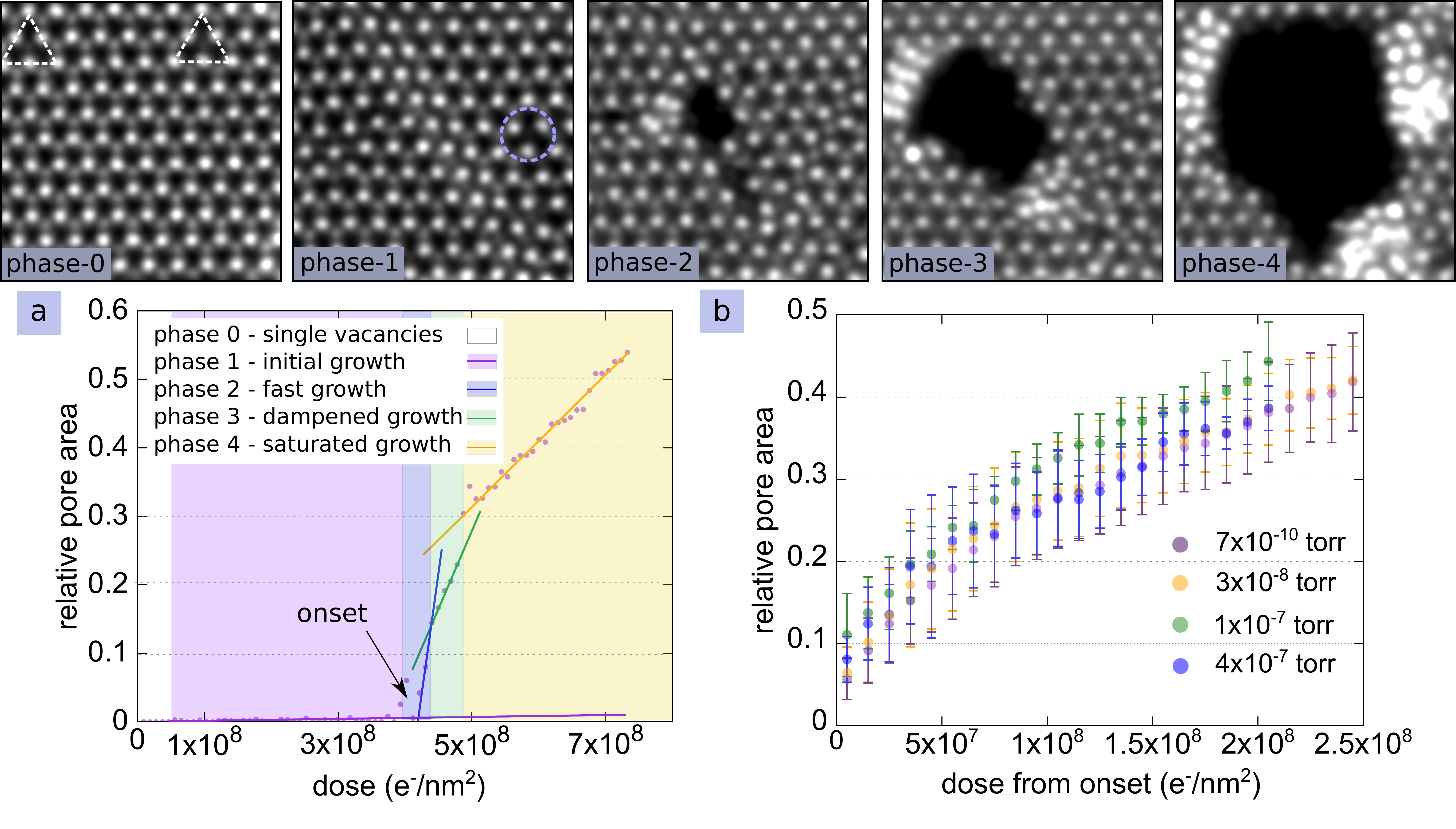}
    \caption {{\bf Pore growth in MoS$_2$.}
    STEM-HAADF images with a field of view of ca. $4\times 4$~nm$^2$ (512$\times$512~pixels), filtered with a Gaussian blur (with a radius of $2$ pixels).
    White triangles highlight single sulfur vacancies, and the dashed blue circle a double vacancy.
    (a) Example of the growth of the pore area as a function of the electron dose recorded at $7\times 10^{-10}$~torr.
    Different phases are highlighted with different colors, and the lines are linear fits to the data to guide the eye.
    (b) Averaged pore areas, starting from the onset marked in (a), for experiments at four different oxygen partial pressures.  
    The error bars correspond to the standard deviations of the data.
    }
\label{fg::growth-example}
\end{center}
\end{figure}

We start the in situ experiments by recording images continuously of the two materials under different oxygen partial pressures.
The base pressure in the objective area of the microscope is ca. $2\times 10^{-10}$~mbar, and oxygen is introduced directly to this volume through a leak valve, as described in Ref.\cite{Leuthner2019}, where it was also estimated that the actual pressure at the sample is approximately one order of magnitude higher than that measured by the objective area pressure gauge.
All pressures given here are readings from the gauge.
In MoS$_2$, continuous imaging practically always resulted in pore formation within the imaged area.
Example images recorded during such an experiment are shown in \textbf{Figure~\ref{fg::growth-example}} recorded with a field of view of approximately $4\times 4$~nm$^2$.
To quantify the pore growth, we measure the pore area in each image as compared to the whole field of view, and plot it as a function of the cumulated electron dose.
An example of one experiment is shown in Figure~\ref{fg::growth-example}a.
Based on the rate of increase in the pore area growth, we divide the pore growth into four consecutive phases ($1-4$) following the initial phase consisting of single vacancies (phase~0).

Phase~1 is initiated by emergence of point defects (typically S vacancies).
Small pores form with the minimum size of a S double vacancy (one example is highlighted with a blue dashed line in the second STEM-HAADF image in Figure~\ref{fg::growth-example}).
Characteristically, sulfur vacancy lines appear at this phase (going from the top left corner to the bottom right in the sample image).
These observations are in line with the literature~\cite{Komsa2013,Ryu2016}.

A single pore appears in phase~2.
Smaller defects merge into the growing pore, as seen in the third STEM-HAADF image in Figure~\ref{fg::growth-example}.
During this phase, dozens of molybdenum and sulfur atoms can be removed during recording a single image corresponding to a dose of about 3$\times$10$^{7}$~e$^{-}/$nm$^2$.
Phase~2 exhibits the fastest recorded growth rate.

In phase~3 pore growth slows down.
Small molybdenum structures start appearing at the pore edges (see the fourth STEM-HAADF image in Figure~\ref{fg::growth-example}), again in line with the literature~\cite{Liu2013,Ryu2016,Huang2018}.
Both molybdenum zigzag (Mo-zz) and sulfur zigzag (S-zz) terminated edges can appear.

Phase~4 displays fully saturated growth with a high proportion of the pore edges covered by molybdenum structures appearing as bright formations (see the last STEM-HAADF image in Figure~\ref{fg::growth-example}).
During this phase, the growth rate is typically only few edge atoms per image.

Data from all experiments for MoS$_2$ are collected in Figure~\ref{fg::growth-example}b.
Each data point is an average of ca. $10 - 20$ measurements with the error bar showing the standard deviation of the data.
As can be seen, the results measured at different oxygen partial pressures overlap and do not show consistent pressure-dependency, indicating that the oxygen atmosphere has a negligible role in the observed damage.
Thus, we conclude that the etching of MoS$_2$ is not enhanced chemically by the oxygen radicals landing on the surface, but is instead driven by the knock-on process (with contribution from inelastic excitations~\cite{Kretschmer2020}) caused by the electron beam.

After the experiments with MoS$_2$, we initially repeated the measurements in a similar manner with MoTe$_2$.
However, this resulted in significant damage {\it outside} the field of view that was used during the experiment. The behaviour is in stark contrast to MoS$_2$ where this was never observed.
Instead, it is similar to what we have observed earlier for graphene~\cite{Leuthner2019}; atomic oxygen created by the electron beam from the O$_2$ molecules in the sample atmosphere led to chemical etching of graphene starting at point defects in the immediate vicinity, but outside the imaged area that only contained pristine graphene.

In order to quantify this process, we changed the experimental procedure to record continuous series with a dynamic imaging area: the field of view was kept at (nominal) $4\times4$~nm$^2$ with one image with a field of view of (nominal) $20\times 20$~nm$^2$ recorded once a minute to allow quantifying the damage of the surrounding area, while simultaneously minimising the electron dose affecting it.
We recorded $15$ to $25$ series at each oxygen partial pressure.
Within the smaller field of view, damage in MoTe$_2$ was similar to what we observed for MoS$_2$ up to phase~2.
However, we did not observe the dampened pore growth due to the lack of Mo-rich edge termination, presumably arising from the higher mass of Te as compared to S, which hinders knock-on displacement of Te at the $60$~kV acceleration voltage.

\begin{figure}
\begin{center}
\includegraphics[width=1.0\textwidth]{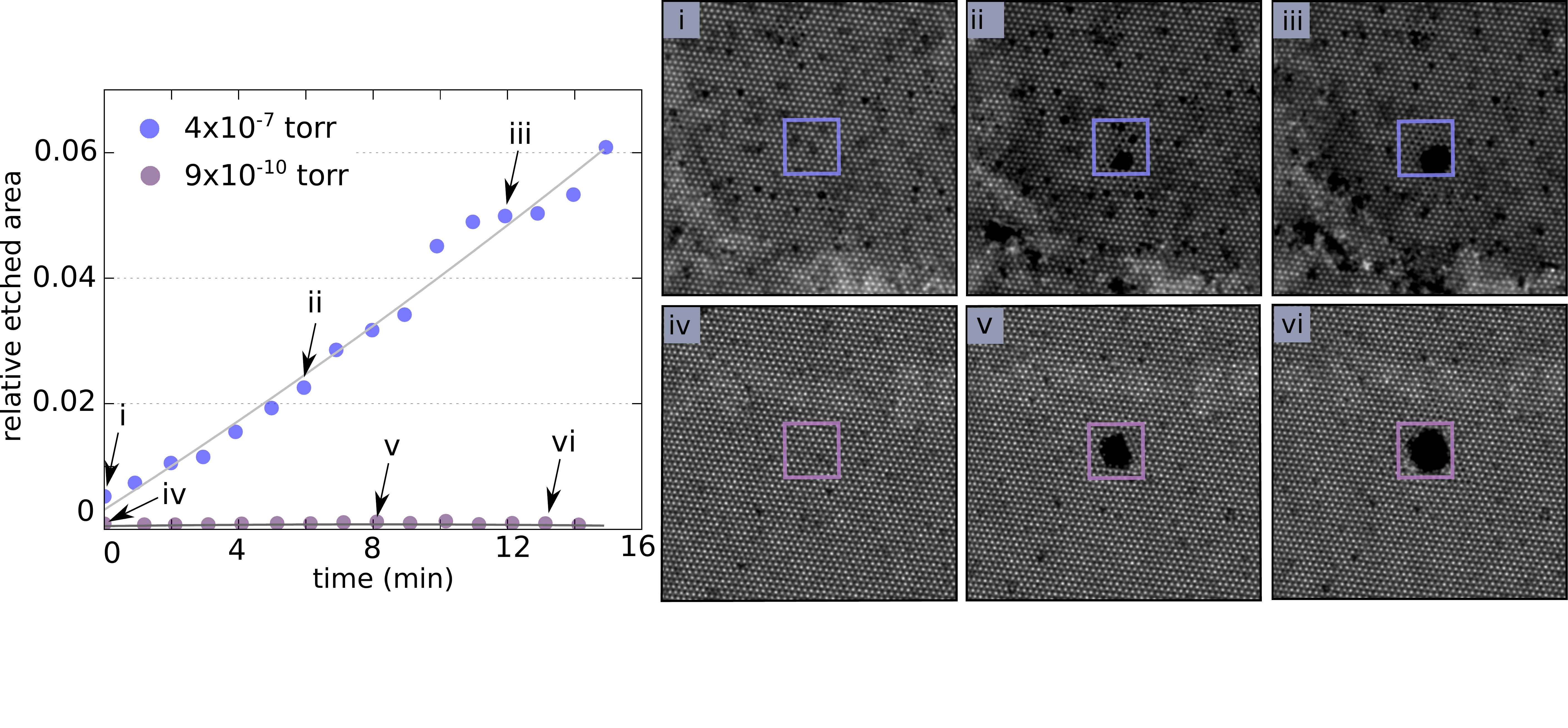}
    \caption {{\bf Chemical etching of MoTe$_2$}.
    Relative etched area outside the smaller field of view is shown for two different oxygen partial pressures as a function of time. Example STEM-HAADF images from both experiments are shown with the smaller field of view excluded from the analysis highlighted with a square. The labels of the images correspond to the data points marked on the plot. The (nominal) field of view for all images is $20\times 20$~nm$^2$. The images have $1024\times1024$ pixels, and they have been filtered with a Gaussian blur (with a radius of 2 pixels).
    }
\label{fg::MoTe2-example}
\end{center}
\end{figure}

To quantify the oxygen-related damage in MoTe$_2$, we only analyzed the total area of all pores outside the smaller field of view.
Instead of electron dose, which is mostly deposited within the smaller field of view, the relative etched area was recorded as a function of time, since the chemical etching process takes place also without the electron beam irradiating the etched area. Each series was up to $12-15$ min long.
Example images and the calculated etched area as a function of time are shown in \textbf{Figure~\ref{fg::MoTe2-example}}.

\begin{figure}
\begin{center}
\includegraphics[width=0.95\textwidth]{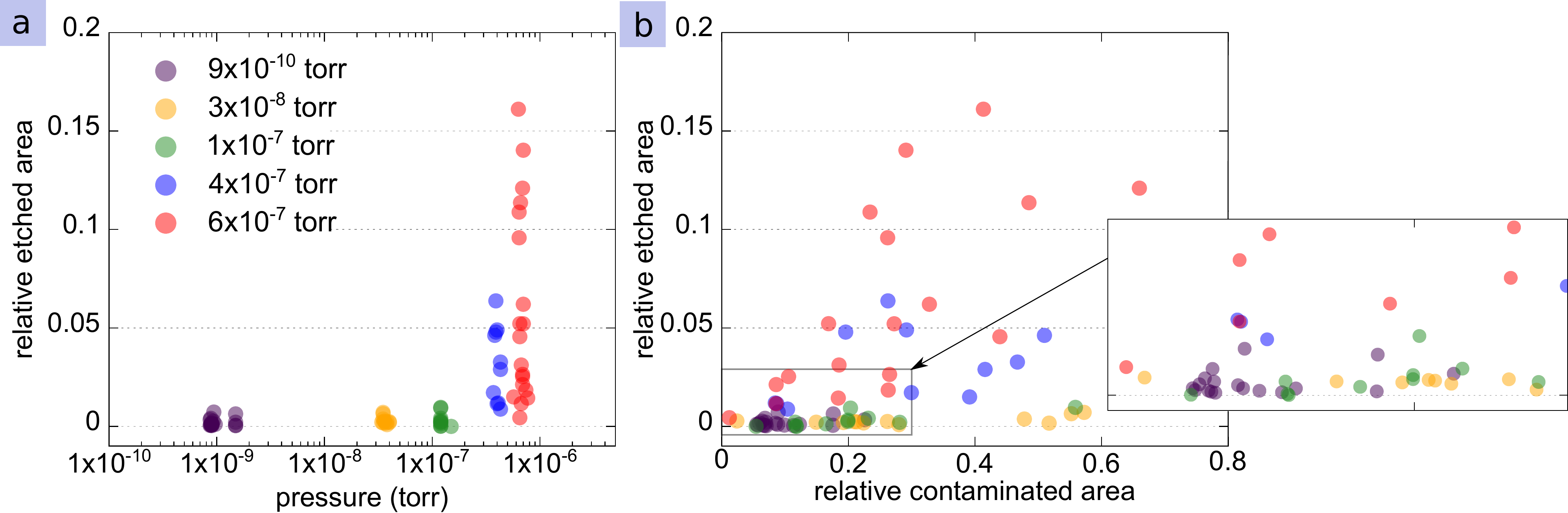}
    \caption {{\bf Relative etched area of MoTe$_2$ as a function of oxygen partial pressure and amount of contamination.}
    (a) The relative etched area at 12~min for each recorded data series as a function of pressure.  
    (b) The relative etched area at the same instance shown as a function of the relative contaminated area estimated from each series. A zoom-in of the low contaminated area is given as an inset.
    }
\label{fg::MoTe2-etching}
\end{center}
\end{figure}

The increase in etched area now follows a second order polynomial,
since both new pores are constantly created and the already existing ones simultaneously grow during the experiment.
Due to the spatial variation in the sample conditions (as described below), it was impossible to deduce accurate values for the pore creation or pore etching rates from the data.
Instead, in \textbf{Figure~\ref{fg::MoTe2-etching}}a we plot the relative area etched after twelve minutes in each experiment as a function of the pressure.
The data reveal a clear dependency between the etched area and the oxygen partial pressure highlighting the role of the oxygen in the observed damage.
However, the large scatter of the data, especially at higher pressures, indicates that there is another factor influencing the damage that either enhances or hinders the etching process.

The most obvious feature that varies between different samples and sample areas in our experiments is the amount of hydrocarbon-based surface contamination. It is common on all surfaces, but particularly important in the case of 2D materials~\cite{Algara-Siller2014,Peng2017}.
To quantify its influence on the etched area, we plot the etched area as a function of the relative contaminated area in the larger field of view  for each of the MoTe$_2$ experiments, shown in Figure~\ref{fg::MoTe2-etching}b.
At lower pressures, where the etching remains suppressed, we don't see a clear trend in the data.
This however clearly changes for the higher pressures, starting at $1\times 10^{-7}$~torr, after which the relative etched area grows nearly linearly as a function of the amount of contamination.
As was reported for graphene~\cite{Leuthner2019}, hydrocarbon contamination naturally contains some oxygen, which can contribute to chemical etching under electron irradiation.
Therefore, we assume that for the experiments here, the contamination serves as a reserve for oxygen which is replenished from the oxygen atmosphere at higher pressures, resulting in increased concentration of oxygen on the sample surface, which leads to the observed dependency.

\begin{figure}
\begin{center}
\includegraphics[width=0.4\textwidth]{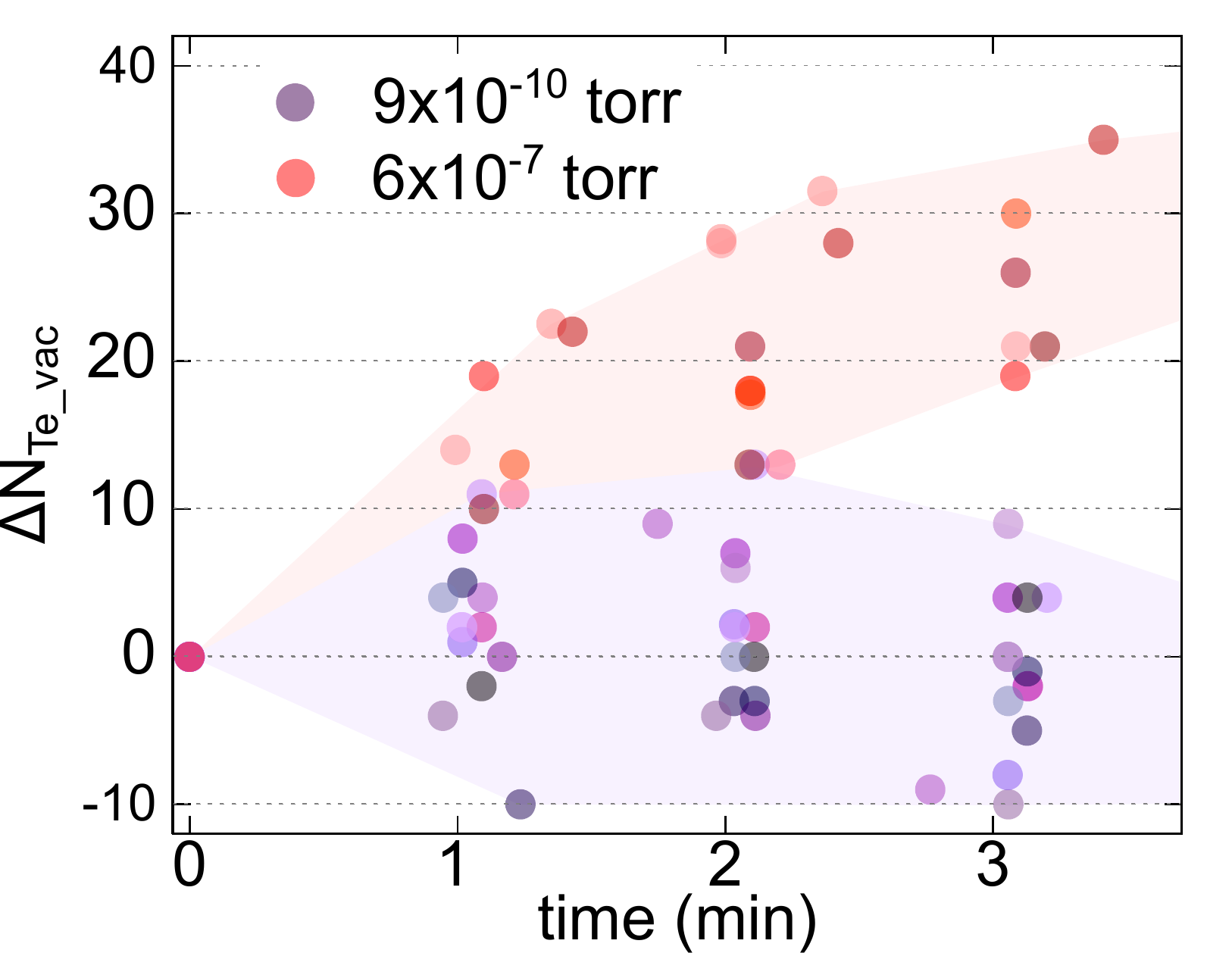}
    \caption {{\bf Relative number of Te vacancies as a function of time.}
    Data for the lowest and the highest pressure experiments are shown as calculated for the first four larger field of view frames of each experiment, with the number of vacancies in the first frame set to zero.
    Each experiment is shown with a unique color.
    The colored background areas are guides to the eye.
    }
\label{fg::MoTe2-sv}
\end{center}
\end{figure}

Notably, in contrast to both MoS$_2$ and graphene, in MoTe$_2$ the damage outside the smaller field of view first appears as individual vacancies also in the clean sample area, that subsequently grow into pores.
It is likely that the process responsible for their formation is related to the oxidation damage process that takes place under ambient conditions.
In \textbf{Figure~\ref{fg::MoTe2-sv}}, we plot the change in the number of Te vacancies $\Delta N_\mathrm{Te_{vac}}$ as a function of time for two different pressures for the four first recorded frames of each experiment (after this, vacancies start to merge and grow into pores).
In ultra high vacuum ($9\times 10^{-10}$~torr), the number of vacancies does not significantly change (the variation in the values is within the accuracy of determining them based on STEM-HAADF image contrast), whereas at the higher pressure ($6\times 10^{-7}$~torr) a clear increase is observed.
Therefore, the reason for vacancy formation must be related to atomic oxygen that is created from the oxygen atmosphere by the electron beam.

In order to estimate how much oxygen is present at the sample surface at each pressure, we calculate the number of oxygen molecules hitting the sample surface per time using the impingement rate $J=PN_A/\sqrt{2\pi MRT}$, where $P$ is the gas pressure, $N_A$ is the Avogadro's number, $M$ is the molecular mass of the gas particle, $R$ is the gas constant and $T$ is the temperature.
The values of $J$ are given in \textbf{Table~\ref{tab:impingement}} for each pressure (calculated with $M_{\mathrm{O}_2}=2\times 15.999$~g/mol and $T=273$~K).
Multiplying the rate $J$ with the imaged area, we get the number of atoms hitting the surface per second (multiplied by two to take into account the two available surfaces in a freestanding mono-layer).
At the highest pressure 1559 oxygen molecules reach the imaged area every second.
Time between each recorded frame, from which the etched area is calculated, is approximately one minute, indicating that up to 93540 oxygen molecules can contribute to the etching at each frame.
The value presents an upper limit for the number of molecules on the surface, because it does not take into account desorption from the surface.
In comparison, the number of molecules available on the surface is significantly lower at the three lowest oxygen partial pressures ($2$, $78$ and $260$ molecules per second corresponds to $120$, $4680$, $15600$ molecules per frame).
At these pressures we do not observe statistically significant etching, implying that the abundance of oxygen available on the surface plays an important role in the etching process.
We explore this further in the computational analysis below.

Previous theoretical studies on oxidation of TMDs have shown that pristine mono-layers remain intact when exposed to O$_2$~\cite{liu_oxidation_2015, farigliano_oxidative_2021, wang_oxidation_2021}, which is also in line with our experimental observations.
However, O$_2$ molecules can readily adsorb on chalcogen vacancy defects where they can dissociate into oxygen radicals.
The dissociation barrier of O$_2$ on single vacancies of MoS$_2$ and MoTe$_2$ were predicted to be about $0.9$~eV and $0.25$~eV respectively~\cite{liu_oxidation_2015, farigliano_oxidative_2021, wang_oxidation_2021}.
It has been shown that such dissociation is an intermediate step for the oxidative etching of MoS$_2$, which, however, is associated with relatively high energy barriers and is not feasible at room temperature~\cite{farigliano_oxidative_2021}.
In our experiments we introduce oxygen radicals directly due to the splitting of O$_2$ under the influence of the electron beam.
However, even in this case we do not observe vacancy formation in MoS$_2$ outside the area directly under electron irradiation.

Our calculated adsorption energies, $E_\mathrm{ads}$, of oxygen radicals on pristine MoS$_2$ and MoTe$_2$ surfaces and on a chalcogen vacancy are shown in \textbf{Table~\ref{tab:adsorp}}.
The results agree well with previously reported values~\cite{he_apl_2010, ma_PCCP_2011, Wu2020}.
In all cases the adsorbed oxygen has a singlet ground state configuration.
For the pristine surfaces the most favourable adsorption configuration of an oxygen radical is on top of a chalcogen atom.
The oxygen adsorption energy on the pristine MoS$_2$ surface was calculated to be $-3.59$~eV and is $0.48$~eV lower compared to the pristine MoTe$_2$ surface.
The adsorption energies of O radicals on a chalcogen vacancy are significantly lower and, in contrast to the pristine surface, the calculated $E_\mathrm{ads}$ of $-7.02$~eV for S vacancy is slightly higher than that for Te vacancy of $-7.18$~eV.

\begin{figure}
\begin{center}
\includegraphics[width=0.9\textwidth]{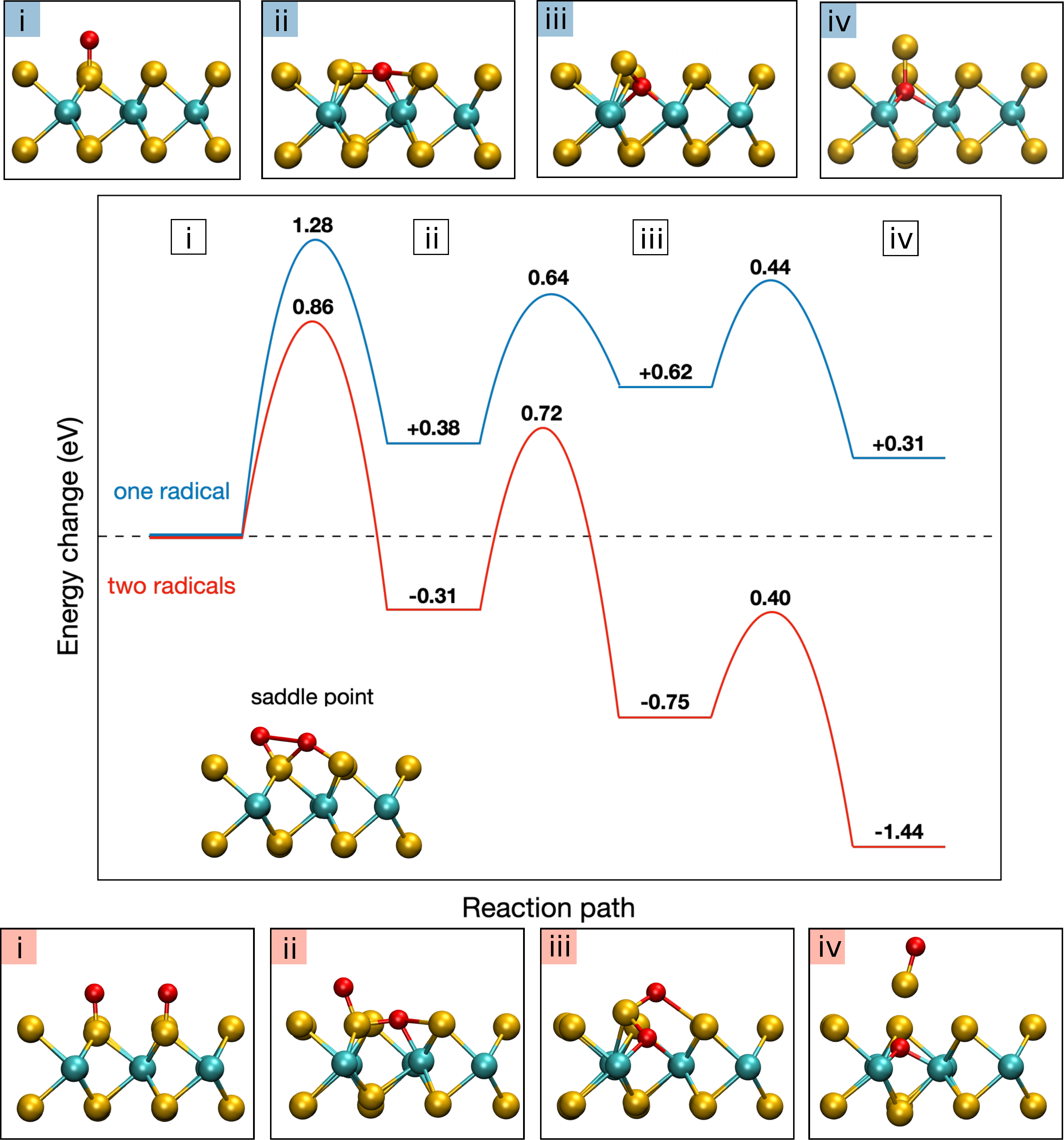}
    \caption {{\bf Reaction path for O intercalation of MoTe$_2$.}
    Schematic energy profile along the reaction path obtained from CI-NEB calculations for intercalation of an oxygen radical (red spheres) between Te (yellow) and Mo (cyan) atoms in MoTe$_2$.
    Initial, final and two intermediate configurations from i) to iv) are shown at the top and bottom panels for one and two oxygen radicals, respectively.
    The inset shows the saddle point configuration for the reaction from i) to ii).
    The numbers with +/- signs show the difference in the system total energy with respect to the initial configuration, while the numbers without a sign correspond to the values of the energy barriers.
    }
\label{fg::DFT_energetics}
\end{center}
\end{figure}

To reveal a possible mechanism for vacancy creation in MoTe$_2$ we studied intercalation of O radicals between Te and Mo atoms using the climbing image nudged elastic band (CI-NEB) method.
We started from the configuration with a single O radical adsorbed on top of a Te atom.
The change in total energy of the system along the reaction path is shown in \textbf{Figure~\ref{fg::DFT_energetics}} in blue.
There are two local minimum configurations: in the first one the oxygen atom adopts a bridge-like structure between Te and Mo atoms, and in the second one it is bound to two Mo atoms and a Te atom between them (structures ii and iii in the top panel of Figure~\ref{fg::DFT_energetics}).
These configurations, however, are higher in energy than the initial one.
The energy barrier between the initial and the first local minimum configurations is 1.28~eV, suggesting that this reaction is kinetically not feasible at room temperature.

The energetics of such process, however, drastically change when there are two oxygen radicals adsorbed on two neighboring Te atoms.
The process becomes exothermic by $1.44$~eV, and, while there are similar local minimum configurations, they are lower in energy than the initial structure with two O radicals adsorbed on top of Te atoms, see the path depicted in red in Figure~\ref{fg::DFT_energetics}.
Moreover, the energy barrier for an O atom to move into the first local minimum drops to $0.86$~eV.
The binding energy of a TeO pair on the oxygen passivated single vacancy is calculated to be only $0.21$~eV, suggesting that it can easily desorb from the surface.
The energy barrier of $0.86$~eV is still relatively high and barely at the limit for the process that can take place at room temperature in a reasonable time.
However, our results demonstrate that etching of MoTe$_2$ by oxygen radicals can occur even for a non-defective mono-layer. The results also support our earlier statement, that the abundance of oxygen plays an important role in facilitating chemical etching in MoTe$_2$.

\begin{figure}
\begin{center}
\includegraphics[width=0.4\textwidth]{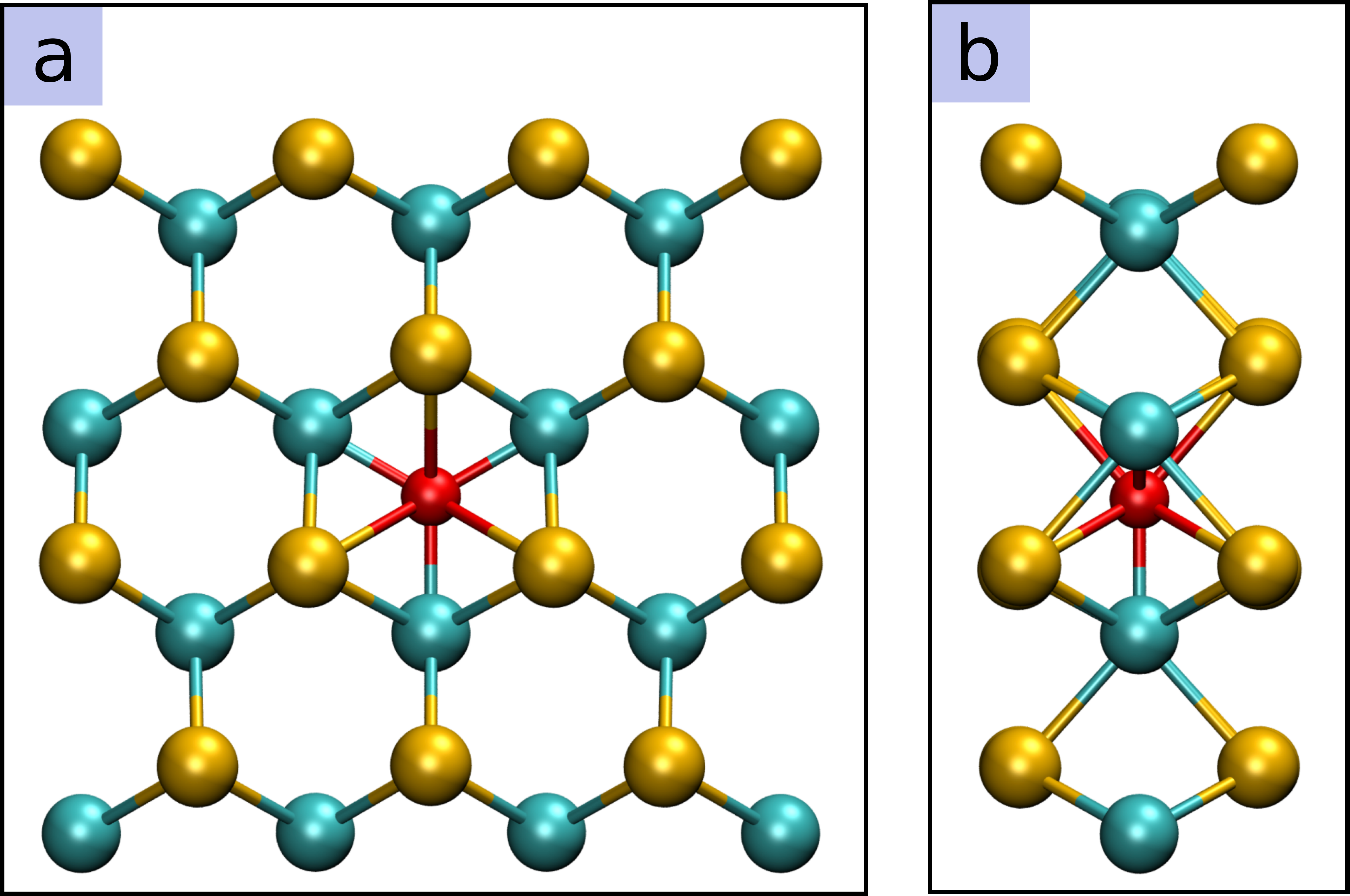}
    \caption {{\bf Interstitial oxygen in MoTe$_2$.}
    Top (a) and side (b) views of the atomic structure of MoTe$_2$ with an oxygen radical positioned in the center of a triangle formed by Mo atoms.
    }
\label{fg::structure1}
\end{center}
\end{figure}

We also found that for MoTe$_2$ the position of an oxygen radical in the center of a triangle formed by Mo atoms (see \textbf{Figure~\ref{fg::structure1}}) is $0.51$~eV more favourable than oxygen adsorbed on top of a Te atom.
In contrast, for MoS$_2$ such configuration is energetically less favourable by $1.75$~eV.
The oxygen atom in this configuration introduces significant lattice strains: the distance between Mo atoms around oxygen is increased by $1.3$\%, while the distance between Te atoms is increased by $5.3$\% and $3.0$\% in the in-plane and out-of-plane directions, respectively.
The binding energy of the surrounding Te atoms (defined in the same way as for oxygen radicals) is decreased by $17$\%, from $4.85$~eV to $4.03$~eV.
Therefore, oxygen atoms in such configuration can facilitate etching and degradation of MoTe$_2$ layers.

CI-NEB calculations suggest that the reaction path and energy profile for an O radical to move into this interstitial configuration from the surface is similar to that shown in Figure~\ref{fg::DFT_energetics} for the reactions from i to iii.
Therefore, such reaction with the energy barrier of $1.28$~eV is not feasible in pristine MoTe$_2$ at room temperature.
In the case of two oxygen radicals, intercalation of one of them between Te and Mo atoms, as shown in Figure~\ref{fg::DFT_energetics},  is energetically more favourable.
However, at the defective sites such configuration can be potentially reached more easily.
For example, when an oxygen molecule dissociates on a Te vacancy, one of the formed O radicals can adopt a position in-between Mo atoms.

Taking into account our experimental findings, which show that surface contamination significantly affects MoTe$_2$ degradation, we have extended our computational analysis to assess the possible effects of the presence of atomic carbon in the oxidative etching mechanism.
We first calculated adsorption energies of C in different configurations defined as follows: Ch – above a chalcogen atom, M – above a metal atom, H – above a triangle formed by Te atoms, ICh – interstitial configuration below a Te atom (as shown for oxygen in the Fig.~\ref{fg::DFT_energetics}, top panel iv), IH - interstitial configuration in the center of a triangle formed by Mo atoms (as shown for oxygen in the Fig.~\ref{fg::structure1}).
The obtained results are summarized in the \textbf{Table~\ref{tab:carbon}}.
It is found that for atomic carbon the interstitial configurations are energetically much more favourable.
Moreover, according to our CI-NEB calculations, the energy barriers for the reaction paths from the M site to the ICh and IH sites are only 0.43 and 0.57 eV, respectively.
These results suggest that in MoTe$_2$ atomic carbon will adopt interstitial configurations.

\begin{figure}
\begin{center}
\includegraphics[width=0.8\textwidth]{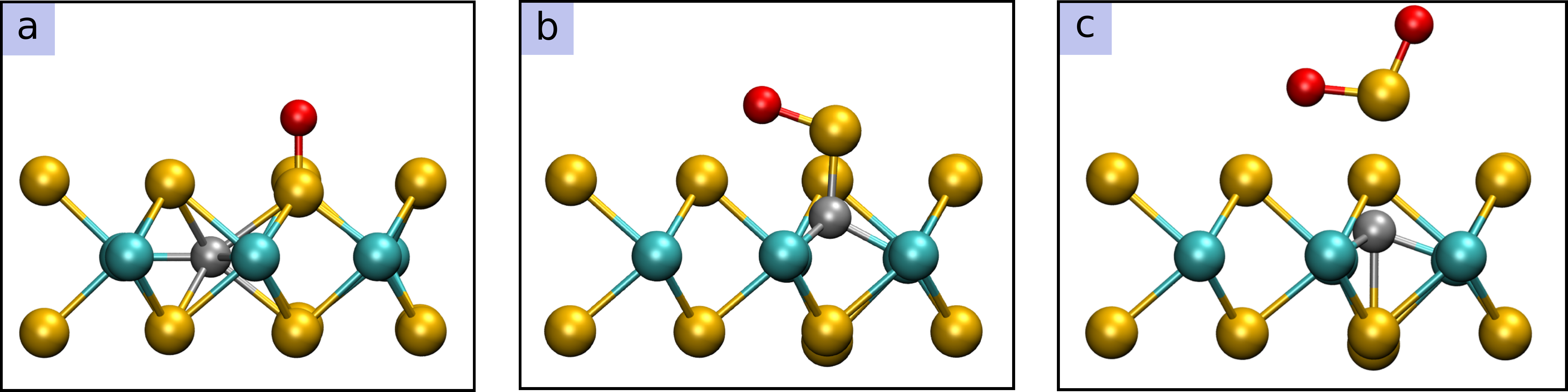}
    \caption {{\bf The effect of carbon on the oxygen etching of MoTe$_2$.} The optimised configurations of adsorbed oxygen (red) when carbon (grey) is intercalated within the structure. a) Oxygen adsorbed on top of chalcogen with carbon interstitial in the IH site at the center of three molybdenum atoms. b) Carbon interstitial in the ICh site below a Te atom. c) Formation of TeO$_2$, weakly bound to the C-passivated chalcogen vacancy, after adsorption of the second oxygen radical. The adsorption energy of TeO$_2$ is only $-0.31$~eV. 
    }
\label{fg::carboxy}
\end{center}
\end{figure}

Further, our calculations show that in the presence of an oxygen radical adsorbed on top of a Te atom, the ICh configuration for carbon (\textbf{Figure~\ref{fg::carboxy}}b) becomes $1.17$~eV more energetically favourable than the IH configuration (Figure~\ref{fg::carboxy}a).
The adsorption energy of a TeO pair on the C-passivated chalcogen vacancy was calculated to be $-1.42$~eV. However, if the second O radical is adsorbed on the same Te atom, the adsorption energy of the resulting TeO$_2$ (Figure~\ref{fg::carboxy}c) is only $-0.31$~eV.
Overall, our results show that atomic carbon preferentially adopts interstitial configurations in MoTe$_2$, which can significantly affect its properties, as well as defect formation and oxidation mechanisms.



\section{Conclusion}

We have experimentally compared oxygen-related chemical damage in mono-layer MoS$_2$ and MoTe$_2$ in situ in varying oxygen atmospheres during transmission electron microscopy at room temperature.
While structural changes in MoS$_2$ show no dependency on the oxygen partial pressure, and remain limited to the area directly under electron irradiation, for MoTe$_2$ significant damage occurs also in the vicinity of the electron probe outside the imaged area.
The damage in MoTe$_2$ increases with increasing oxygen partial pressure (starting at ca. $1\times 10^{-7}$~torr) revealing its chemical nature, and within experiments at the same pressure is pronounced in areas with more hydrocarbon-based surface contamination.
In ultra high vacuum, both materials damage similarly with the exception of increasing Mo concentration in MoS$_2$, presumably caused by knock-on displacement of sulfur atoms.
We report significant etching in MoTe$_2$ above the pressure of 1$\times$10$^{-7}$ torr.
At pressures below this, the etching is statistically not significant and we even see healing of point defects induced by the electron beam.
Importantly, the chemically driven damage in MoTe$_2$ starts as individual Te vacancies also in the clean sample area, which is in contrast to both graphene and MoS$_2$.
To reveal the underlying mechanism, we carried out ab initio simulations, which show that two neighboring oxygen atoms on top of Te atoms in MoTe$_2$ enable an energetically favorable reaction pathway for oxygen intercalation that leaves a weakly bound ($0.21$~eV) TeO pair on top of the surface that can desorb even at room temperature.
Overall, the results demonstrate clearly the significant difference between MoS$_2$ and MoTe$_2$ under oxygen revealing also the underlying atomic-scale processes. 


\section{Methods Section}

The MoS$_2$ samples were grown by chemical vapour deposition~\cite{O'Brien2014} on SiO$_2$ and the MoTe$_2$ samples mechanically exfoliated, after which both were transferred onto a supported gold grid with holey carbon membrane (Quantifoil).
Samples were baked in vacuum overnight at ca. 150$^{\circ}$C before being inserted into the scanning transmission electron microscopy instrument Nion UltraSTEM 100.
The microscope was operated at a $60$~kV accelerating voltage with the convergence semi-angle of the electron probe of $30$~mrad.
Images were recorded both in bright field at lower resolution with the Ronchigram camera as well as at atomic resolution with a high angle annular dark field (HAADF) detector with collection angles of $80–300$~mrad.
The operation of the leakvalve system coupled to the microscope followed the method described in~\cite{Leuthner2019}.
The STEM-HAADF images were processed with Gaussian filter to reduce noise and enhance contrast before analysing the images. 

The density functional theory calculations were conducted with the GPAW package~\cite{enkovaara_electronic_2010} using the Perdew-Burke-Ernzerhof (PBE) exchange-correlation functional~\cite{perdew_generalized_1996} and a plane-wave basis set with an energy cutoff of $600$~eV.
Mono-layers of MoS$_2$ and MoTe$_2$ were modelled using $4\times4\times1$ supercells and a vacuum of 20~{\AA} to separate slabs in the $z$-direction. 
The Brillouin zone was sampled using a $4\times4\times1$ \textbf{k}-point mesh according to the Monkhorst-Pack scheme~\cite{monkhorst_special_1976}. 
The smearing of occupation numbers was determined using a Fermi-Dirac distribution with a width of 0.025~eV and spin-polarization was taken into account.
Transition states and energy barriers were obtained using the climbing-image nudged elastic band method (CI-NEB)~\cite{henkelman_climbing_2000} with at least five images in-between initial and final configurations.
For structural optimisations and CI-NEB calculations the convergence criterium of 0.02~eV/\r{A} for the forces was adopted. 

The adsorption energy of an oxygen radical on the pristine TMD's surface and a chalcogen vacancy was calculated as $E_\mathrm{ads} = E_\mathrm{tot}(\mathrm{MoX}_2-\mathrm{O}) - E_\mathrm{tot}(\mathrm{O}) - E_\mathrm{tot}(\mathrm{ MoX}_2)$, where $E_\mathrm{tot}(\mathrm{MoX}_2-\mathrm{O})$ is the total energy of the TMD with the adsorbed oxygen, $E_\mathrm{tot}(\mathrm{O})$ is the energy of the oxygen radical and $E_\mathrm{tot}(\mathrm{MoX}_2)$ is the total energy of the TMD with or without a chalcogen vacancy.
We note that the definition of the adsorpton energy can vary in the literature.
Here we used the calculated energy of a separate O atom in the triplet configuration as a reference energy for oxygen, while in some articles the chemical potential, defined as half the energy of an oxygen molecule, is used instead.
Within our methodology, the energy difference between a separate O atom and a chemical potential is $3.13$~eV.

\medskip

\medskip
\textbf{Acknowledgements} \par 
EH{\AA}, AM and SS acknowledge funding from Austrian Science Fund (FWF) through project number M2595. Computational resources from the Vienna Scientific Cluster (VSC) are gratefully acknowledged.

\medskip

%

\nocite{phaidra}
\bibliographystyle{ieeetr}
\bibliography{refs}

\begin{table}
\caption{\label{tab:impingement} 
    {\bf Number of oxygen molecules impinging the MoTe$_2$ surface per time.}
    The calculated impingement rate $J$ and the number of oxygen molecules reaching the sample surface area per time at each oxygen partial pressure are calculated as described in the text.
    }
\begin{tabular}{lcc}
\hline
\textrm{Pressure (torr)}&
\textrm{$J$ (m$^{-2}$s$^{-1}$)}&
\textrm{molecules per time (s$^{-1}$)}\\
\hline
9 $\times 10^{-10}$ & 3.38$\times 10^{15}$ & 2 \\
3 $\times10^{-8} $ & 1.13$\times 10^{17}$ & 78 \\
1 $\times10^{-7} $ & 3.76$\times 10^{17}$ & 260 \\
4 $\times10^{-7} $ & 1.50$\times 10^{18}$ & 1039 \\
6 $\times10^{-7} $ & 2.26$\times 10^{18}$ & 1559\\

\hline
\end{tabular}
\end{table}

\begin{table}
\caption{\label{tab:adsorp} 
    {\bf Oxygen adsorption energies.}
    Calculated adsorption energies, $E_\mathrm{ads}$, in eV of an oxygen radical (O) on the pristine MoS$_2$ and MoTe$_2$ surfaces and on a chalcogen (X) vacancy.
    The values obtained with both the plane wave (PW) and a linear combination of atomic orbitals (LCAO, $dzp$ basis set) description of the wave function are given for comparison. 
    }
\begin{tabular}{lcc}
\hline
\textrm{System}&
\textrm{Pristine PW/LCAO}&
\textrm{X-vacancy PW/LCAO}\\
\hline
MoS$_2$  & -3.59/-3.34 & -7.02/-6.98  \\
MoTe$_2$ & -3.11/-2.78 & -7.18/-7.26 \\
\hline
\end{tabular}
\end{table}

\begin{table}
\caption{\label{tab:carbon} 
    {\bf Carbon adsorption energies.}
    Calculated adsorption energies ($E_\mathrm{ads}$) and magnetic moments ($M$) of a carbon atom in different configurations in MoTe$_2$ defined as follows: Ch – above a chalcogen atom, M – above a metal atom, H – above a triangle formed by Te atoms, ICh – interstitial configuration below a Te atom (as shown for oxygen in the Fig.~\ref{fg::DFT_energetics}, top panel iv), IH - interstitial configuration in the center of a triangle formed by Mo atoms (as shown for oxygen in the Fig.~\ref{fg::structure1}). 
    }
\begin{tabular}{lcc}
\hline
\textrm{Ads. site}&
\textrm{$E_\mathrm{ads}$ (eV)}&
\textrm{$M$ ($\mu_B$)}\\
\hline
Ch  & -1.70 & 2.0  \\
M   & -3.18 & 1.9  \\
H   & -1.57 & 2.0  \\
ICh & -5.03 & 0.0  \\
IH  & -5.21 & 0.0  \\
\hline
\end{tabular}
\end{table}

\begin{figure}
\textbf{Table of Contents}\\
\medskip
  \includegraphics{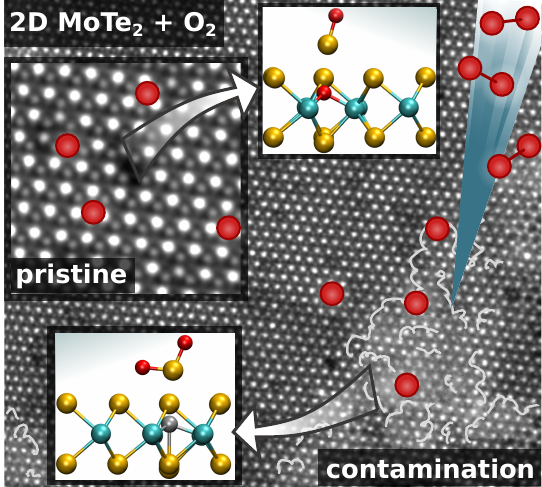}
  \medskip
  \caption*{Oxidation leads to the degradation of 2D MoTe$_2$ under in situ electron microscopy at low oxygen pressures, while 2D MoS$_2$ surface remains inert. The etching is facilitated by abundant oxygen at partial oxygen pressures above 1$\times$10$^{-7}$~torr. The atomic scale etching mechanism is revealed computationally. Hydrocarbon contamination, commonly found on surfaces, accelerates the etching process over forty times.}
\end{figure}

\end{document}